\documentclass[12pt]{iopart}
\usepackage{amssymb}
\usepackage{graphicx}	% Including figure files
\usepackage{url} % provides @online ou \url for bibtex
\usepackage{hyperref} % Hyperrefs and pdf settings
\usepackage{xcolor}

\newcommand{\Msun}{{\rm M}_\odot} %nominal solar mass unit (cf. IAU 2015 resolution B3)

\begin{document}

\title[Closing a spontaneous-scalarization window with binary pulsars]{Closing a spontaneous-scalarization window with binary pulsars}

\author[Junjie Zhao et al.]{Junjie~Zhao$^{1,2}$, Paulo C. C. Freire$^{3}$, Michael Kramer$^{3,4}$, Lijing~Shao$^{5,3,6}$, Norbert Wex$^{3}$}

\address{$^{1}$ School of Physics and State Key Laboratory of Nuclear Physics
and Technology, Peking University, Beijing 100871, China} 
\address{$^{2}$ Department of Astronomy, Beijing Normal University, Beijing 100875, China}
\address{$^{3}$ Max-Planck-Institut f\"ur Radioastronomie, Auf dem H\"ugel 69,
D-53121 Bonn, Germany}
\address{$^{4}$ Jodrell Bank Centre for Astrophysics, School of Physics and Astronomy,
The University of Manchester, M13 9PL, UK}
\address{$^{5}$ Kavli Institute for Astronomy and Astrophysics, Peking
University, Beijing 100871, China} 
\address{$^{6}$ National Astronomical Observatories, Chinese Academy of
Sciences, Beijing 100012, China}

\ead{junjiezhao@pku.edu.cn {\rm (JZ)}; lshao@pku.edu.cn {\rm (LS)}}

\begin{abstract}
Benefitting from the unequaled precision of the pulsar timing technique, binary pulsars
are important testbeds of gravity theories, providing some of the
tightest bounds on alternative theories of gravity. One class of well-motivated
alternative gravity theories, the scalar-tensor gravity, predict large deviations from general
relativity for neutron stars through a nonperturbative phenomenon known as
spontaneous scalarization. This effect, which cannot be tested in the Solar System,
can now be tightly constrained using the latest results from the timing of a set of
7 binary pulsars (PSRs J0348+0432, J0737$-$3039A, J1012+5307, J1738+0333,
J1909$-$3744, J1913+1102, and J2222$-$0137), especially with the updated parameters of PSRs
J0737$-$3039A, J1913+1102, and J2222$-$0137. Using new timing results, we constrain the neutron
star's effective scalar coupling, which describes how strongly neutron stars couple to
the scalar field, to a level of $\left|\alpha_{\rm A}\right| \lesssim 6 \times 10^{-3}$  in a Bayesian analysis.
Our analysis is thorough, in the sense that our results apply
to all neutron star masses and all reasonable equations of state of dense matters,
in the full relevant parameter space. It excludes  
the possibility of spontaneous scalarization of neutron stars, at least within a class of scalar-tensor gravity theories.
\end{abstract}

\section{Introduction}

In Einstein's theory of general relativity (GR)~\cite{Einstein:1915ca},
gravitational waves (GWs) have a quadrupolar structure and, to the leading order,
binary systems lose their orbital energy and angular momentum via GWs according
to the quadrupole formula~\cite{Einstein:1918btx,Peters:1964zz}. This has been
verified experimentally since the detection of orbital decay
caused by the emission of GWs in the radio pulsar--neutron star (PSR-NS)
system PSR B1913+16, the famous Hulse-Taylor binary discovered in
1974~\cite{Hulse:1974eb,Taylor:1979zz}. Indeed, the observed orbital decay
matches the GR expectation for the emission of quadrupolar waves within
experimental precision~\cite{Weisberg:2016jye}; for a review, see
Ref.~\cite{Damour:2014tpa}.

However, discoveries like cosmological inflation, dark energy, and 
dark matter have greatly motivated the development of many alternative theories 
of gravity~\cite{Fujii:2003pa}. Among the best studied---and among the few that
can make testable predictions for binary pulsars---are the 
$T_1(\alpha_0,\beta_0)$ family of mono-scalar-tensor theories of gravity, proposed by
Damour and Esposito-Far\`ese \cite{Damour:1993hw} (hereafter, the DEF gravity).
These include the Jordan-Fierz-Brans-Dicke
gravity~\cite{Jordan:1949zz, Jordan:1959eg, Fierz:1956zz, Brans:1961sx} as the special
sub-class with $\beta_0 = 0$. These and other scalar-tensor gravity (STG) theories 
predict that, in addition to the quadrupolar GWs predicted by GR,
there are additional monopolar and dipolar components to GW emission
arising from scalar fields.  Consequently, binary star systems
consisting of scalarized bodies evolve faster in their orbital decay.  Here the
dipolar GW emission is of particular importance since it enters the
equations of motion of a two-body system already at the $1.5$ post-Newtonian (PN)
level \cite{Mirshekari:2013vb}. In contrast, the contribution of
quadrupolar radiation corresponds to $2.5$\,PN at the lowest order.

In the presence of a scalar field, the amount of dipolar GW emission depends
strongly on the {\em difference} between the effective scalar couplings, denoted as
$\alpha_{\rm A}$ for body ``$A$'', of the binary components. The quantity $\alpha_{\rm A}$ measures the
effective coupling strength between the scalar field and the scalarized star.
Depending on the parameters of the specific STG theory, $\alpha_{\rm A}$ can be of order unity for
scalarized NSs even if its weak-field limit, $\alpha_0$, is very close to zero~\cite{Yagi:2021loe}.
This nonperturbative effect is known as ``spontaneous scalarization''\cite{Damour:1993hw}. Because
$\alpha_0$ is small, such a possibility in the strong-field regime cannot be
tested with Solar System experiments. Recently, Refs.~\cite{Abbott:2018lct,Zhao:2019suc,Niu:2021nic} have
investigated the constraints on dipolar GW emission using GW observations from the coalescences of compact objects~\cite{TheLIGOScientific:2017qsa,LIGOScientific:2020ibl,LIGOScientific:2021qlt}.

The close agreement between the measured orbital decay of the Hulse-Taylor
pulsar and the GR prediction for quadrupolar GW emission had already introduced, in
the mid-1990s, important
constraints on DEF gravity \cite{Damour:1996ke}. However, given the vastly different gravitational
binding energies of NSs and white dwarfs (WDs), it was already clear that in general
they have very different couplings to the scalar field of the theory.
This difference results in a potentially
observable contribution to the orbital period change of PSR-WD systems
arising from the emission of dipolar GWs, or in its absence, very
tight constraints on STG theories \cite{Damour:1996ke, Shao:2017gwu,Anderson:2019eay}.

Indeed, some of the most stringent limits on dipolar GW emission and DEF gravity obtained since
then have been derived from the measurements of orbital decay of PSR-WD systems. Two
prime examples are PSRs~J1738+0333 \cite{Antoniadis:2012vy,Freire:2012mg} and J0348+0432 \cite{Antoniadis:2013pzd}: their observed orbital decays
conform to the GR expectation within experimental precision and no dipolar GW emission is
detected~\cite{Wex:2014nva, Shao:2016ezh}. This also has considerably tightened the constraints on the
existence of spontaneous scalarization.

Given the comparably large difference in NS
masses between PSRs~J1738+0333 and J0348+0432 ($1.47_{-0.06}^{+0.07}\,\Msun$~\cite{Antoniadis:2012vy} and $2.01 \pm
0.04\,\Msun$~\cite{Antoniadis:2013pzd} respectively), their limits on
dipolar GW emission do not exclude spontaneous scalarization entirely: as first pointed out by
Shibata {\it et al.}~\cite{Shibata:2013pra}, for certain equations of state (EOS) for matter at densities above that
of the atomic nucleus,
particularly those that predict a maximum NS mass near $2 \, \Msun$,
spontaneous scalarization can still occur for NS masses in between those of
PSRs~J1738+0333 and J0348+0432. This is the ``{\it mass gap of spontaneous
scalarization}'', which approximately extends in the NS mass interval  $\left[ 1.60 \,\Msun,1.95
\,\Msun \right]$, as suggested by Shao {\it et al.}~\cite{Shao:2017gwu} in 2017.\footnote{For extremely stiff EOSs
spontaneous scalarization
could in principle occur in NS masses above $2\,\Msun$. Such EOSs, however,
have been excluded in the meantime by the binary NS merger event GW170817
\cite{LIGOScientific:2017vwq}.}

To fill this gap, similar radiative tests~\cite{Shao:2017gwu,Zhao:2019suc} have been conducted
for NSs with masses within this mass gap, like the PSR-WD systems,
PSRs~J1012$+$5307~\cite{Lazaridis:2009kq,Antoniadis:2016hxz,Desvignes:2016yex},
J1909$-$3744~\cite{Desvignes:2016yex,Reardon:2015kba,NANOGrav:2017wvv}, and
J2222$-$0137 \cite{Cognard:2017xyr}. Although by 2017 
such systems were already useful, they were not yet as constraining as the tests
for PSRs~J1738+0333 and J0348+0432~\cite{Shao:2017gwu,Zhao:2019suc}, and for that reason
the mass gap of spontaneous scalarization remained open.

As we show in this paper, the situation has now changed because of a set of
new experimental results which we list in Sec.~\ref{sec:new_results}.
In a Bayesian analysis of these results in Sec.~\ref{sec:method}, we find an upper limit in Sec.~\ref{sec:limits}
on the scalar coupling to {\it any} NSs of $\left|\alpha_{\rm A}\right| \lesssim 6 \times 10^{-3}$.
This means that spontaneous scalarization---{\em the} prominent 
non-perturbative deviation from GR predicted by STG theories---is now ruled out
for all NS masses of all existing EOSs at such a level, at least within a class of massless STG theories that we consider.

\section{New experimental results}
\label{sec:new_results}

We now list the recent experimental results of interest to our investigation.
One of the novelties is that two special PSR-NS systems have become useful for our tests,
which were previously dominated by PSR-WD systems.

First, the previous measurement of the
orbital decay of PSR~J1012+5307 is now complemented by improved spectroscopic mass
measurements \cite{MataSanchez:2020pys} and a new, more precise distance measurement
\cite{Ding:2020sig}. These allow the determination of more precise limits on the dipolar GW emission
in this system. 

Second, the parameters for PSR~J1909$-$3744 have also been 
updated~\cite{Liu:2020hkx}.

Third, and more importantly, the masses and the variation of the orbital period of 
PSR~J2222$-$0137 are now measured, respectively, 6 and 12 times more precisely than in the
previous study of this system~\cite{Guo:2021bqa}.

Fourth, measurements of the component masses and the rate of orbital decay for PSR J1913$+$1102,
a recently discovered PSR-NS system~\cite{Lazarus:2016hfu}, are now available~\cite{Ferdman:2020huz}.
This system is important because the pulsar's mass ($1.62 \pm 0.03 \,\Msun$) is within the mass gap.
Furthermore, the NS companion is much lighter, $1.27 \pm 0.03 \,\Msun$, than the visible pulsar.
This mass asymmetry results in a comparably large asymmetry in the compactness ${\cal C}_A
\equiv GM_A/c^2R_A$ of the two components  (${\cal C}_{\rm c} \approx
0.75\,{\cal C}_{\rm p}$), where $R_A$ denotes the radius of body $A$, and subscripts ``p'' and ``c'' refer to the pulsar and its companion respectively. In the
strong gravitational fields of a NS, $\alpha_{\rm A}$ depends in a strongly non-linear
way on ${\cal C}_A$, which can lead to an order-one asymmetry in the $\alpha_{\rm A}$
of the components of this binary and therefore significant dipolar GW emission.

Fifth, new measurements have become available from the timing of PSR~J0737$-$3039A~\cite{Kramer:2021jcw}.
This system is important because the new measurement of its orbital decay is
250 times more precise than the previously published value~\cite{Kramer:2006nb}, and about 25
times more precise than the latest limit on the Hulse-Taylor binary~\cite{Weisberg:2016jye}.
This extreme precision more than compensates for the relatively small mass difference between
the two NSs in the system and the fact that none of them is in the mass range where spontaneous scalarization can still occur, which starts at $1.6\,\Msun$ for most EOSs.

Overall, we collect the relevant updated parameters from these pulsars in three tables: the relevant parameters for the five asymmetric PSR-WD binaries are shown in Tables~\ref{tab:NSWD3s} and~\ref{tab:extraNSWD}, and the relevant parameters from the two PSR-NS systems are listed in Table~\ref{tab:DNS}. The seven binary pulsar systems collected in these three tables densely cover a wide NS-mass range, from 1.25\,$\Msun$ to about 2.0\,$\Msun$.\footnote{It is quite likely that $2\,\Msun$ might be the upper end of the so far measured NS masses. There are arguments that the mass of the current record holder PSR J0740$+$6620 \cite{Fonseca:2021wxt} does not (significantly) exceed $2\,\Msun$ \cite{Miller:2021qha}.}
Quite generally, given the rather large uncertainty in our knowledge of the EOS of NS matter, such a dense coverage is key for constraining certain highly non-linear strong-field deviations from GR, like spontaneous scalarization in DEF gravity.

%--- PSR-WD systems with high-resolution spectroscopy ---

\begin{table}
  \caption{\label{tab:NSWD3s} Relevant binary parameters of the three PSR-WD systems: PSRs
  J0348+0432~\cite{Antoniadis:2013pzd},
  J1012+5307~\cite{Desvignes:2016yex,MataSanchez:2020pys,Ding:2020sig},
and J1738+0333~\cite{Antoniadis:2012vy,Freire:2012mg}. The italic parameters $m_{\rm p}^{\rm obs}$ are
derived from the observed mass ratio $q$ and $m_{\rm c}^{\rm obs}$, obtained from high-resolution spectroscopy observations of the WDs in combination with WD models. The standard 1-$\sigma$ uncertainties are given in the units of the least significant digits in parentheses.}
\begin{indented}
\lineup
\item[]
\begin{tabular}{@{}*{4}{l}}  \br
 Pulsar                                                          & J0348+0432 & J1012+5307 & J1738+0333 \\ \mr
  Orbital period, $P_{\rm b}$\,(d)                                      & 0.102424062722(7)                    & 0.604672722901(13)
  & 0.3547907398724(13)             \\
  Eccentricity, $e$                                               & $2.4(10) \times 10^{-6}$              & $1.30(16)  \times 10^{-6}$                                                               & $3.4(11) \times 10^{-7}$        \\
  Observed $\dot P_{\rm b}$, $\dot P^{\rm obs}_{\rm b}$\,(${\rm fs\,s}^{-1}$) & $-273(45)$                           & $61(4)$                                                                              & $-17.0(31)$                     \\
  Excess $\dot P_{\rm b}$, $\dot P^{\rm xs}_{\rm b}$\,(${\rm fs\,s}^{-1}$)    & $-15(46)$                            & $4.8(51)$                                                                             & $3.15(369)$                     \\
  Mass ratio, $q\equiv m_{\rm p}^{\rm obs} / m_{\rm c}^{\rm obs}$             & 11.70(13)                            & 10.44(11)                                                                             & 8.1(2)                          \\
  Pulsar mass, $m_{\rm p}^{\rm obs}$\,($\Msun$)                     & ${\it  2.01(4)}$                     & ${\it 1.72(16)} $                                                                     & ${\it 1.47(7)}$                 \\
  Companion mass, $m_{\rm c}^{\rm obs}$\,($\Msun$)                  & $0.172(3)$                           & $0.165(15)$                                                                           & $0.181(8)$                      \\ \br
  \end{tabular}
\end{indented}
\end{table}

%--- PSR-WD systems with Shapiro delay ---

\begin{table}
  \caption{\label{tab:extraNSWD} Same as Table~\ref{tab:NSWD3s}, but for two PSR-WD systems where the ``observed'' masses have been obtained from the observed Shapiro delay assuming GR:
  PSRs J1909$-$3744~\cite{Liu:2020hkx} and J2222$-$0137~\cite{Guo:2021bqa}.}
\begin{indented}
\lineup
\item[]\begin{tabular}{@{}*{3}{l}}  \br
  Pulsar                                                          & J1909$-$3744 & J2222$-$0137                    \\ \mr
  Orbital period, $P_{\rm b}$\,(d)                                      & 1.533449474305(5)               & 2.44576437(2)                                       \\
  Eccentricity, $e$                                               & $1.15(7) \times 10^{-7}$        & 0.00038092(1)                                      \\
  Observed $\dot P_{\rm b}$, $\dot P^{\rm obs}_{\rm b}$\,(${\rm fs\,s}^{-1}$) & $-510.87(13)$                   & $250.9(76)$                                       \\
  Excess $\dot P_{\rm b}$, $\dot P^{\rm xs}_{\rm b}$\,(${\rm fs\,s}^{-1}$)    & $-1.7(78)$                      & $-6.3(76)$                                      \\
  Pulsar mass, $m_{\rm p}^{\rm obs}$\,($\Msun$)                     & 1.492(14)                       & 1.831(10)                                       \\
  Companion mass, $m_{\rm c}^{\rm obs}$\,($\Msun$)                  & 0.209(1)                        & $1.3194(40)$                                     \\ \br
  \end{tabular}
\end{indented}
\end{table}

%--- DNS systems ---

\begin{table}
  \caption{\label{tab:DNS} Same as Table~\ref{tab:NSWD3s}, but for two PSR-NS systems:
  PSRs J1913+1102~\cite{Ferdman:2020huz} and
J0737$-$3039A~\cite{Kramer:2021jcw}. The given ``observed'' masses are taken from Refs.~\cite{Ferdman:2020huz} and \cite{Kramer:2021jcw} respectively, where they have been determined from the two most constraining observed post-Keplerian parameters assuming GR.}
\begin{indented}
    \lineup
    \item[]\begin{tabular}{@{}*{3}{l}}  \br
  Pulsar                                                          & J1913+1102 & J0737$-$3039A \\ \mr
  Orbital period, $P_{\rm b}$\,(d)                                      & 0.2062523345(2)                   & 0.1022515592973(10)                              \\
  Eccentricity, $e$                                               & 0.089531(2)                       & 0.087777023(61)                                  \\
  Observed $\dot P_{\rm b}$, $\dot P^{\rm obs}_{\rm b}$\,(${\rm fs\,s}^{-1}$) & $-480(30)$                        & $-1247.920(78)$                                      \\
  Excess $\dot P_{\rm b}$, $\dot P^{\rm xs}_{\rm b}$\,(${\rm fs\,s}^{-1}$)    & $-34.0(285)$                      & $0.05(8)$                                        \\
  Einstein delay parameter, $\gamma^{\rm obs}_{\rm E}$ (ms) & ... & $0.384045(94)$ \\
  Pulsar mass, $m_{\rm p}^{\rm obs}$\,($\Msun$)                     & 1.62(3)                           & 1.33818(1)                                       \\
  Companion mass, $m_{\rm c}^{\rm obs}$\,($\Msun$)                  & 1.27(3)                           & 1.24887(1)                                      \\ \br
  \end{tabular}
\end{indented}
\end{table}

%----------------------------------------------------------------------------------------
\section{The Bayesian inference framework}
\label{sec:method}
%----------------------------------------------------------------------------------------

Here, we investigate the spontaneous scalarization in the DEF theory and obtain
the bounds on scalar couplings $\alpha_{\rm A}$ from the observed binary pulsar parameters
with the help of Bayesian inference. We collect the latest observations of
7 binary pulsars and perform the Markov-chain Monte Carlo (MCMC) technique with EOSs to update the
posterior distributions of the relevant parameters. As the likelihood function
is evaluated over and over, these distributions will converge to the proper and stable
results, which are consistent with astrophysical observations.

\begin{figure*}
  \centering
  \includegraphics[width=0.7\textwidth]{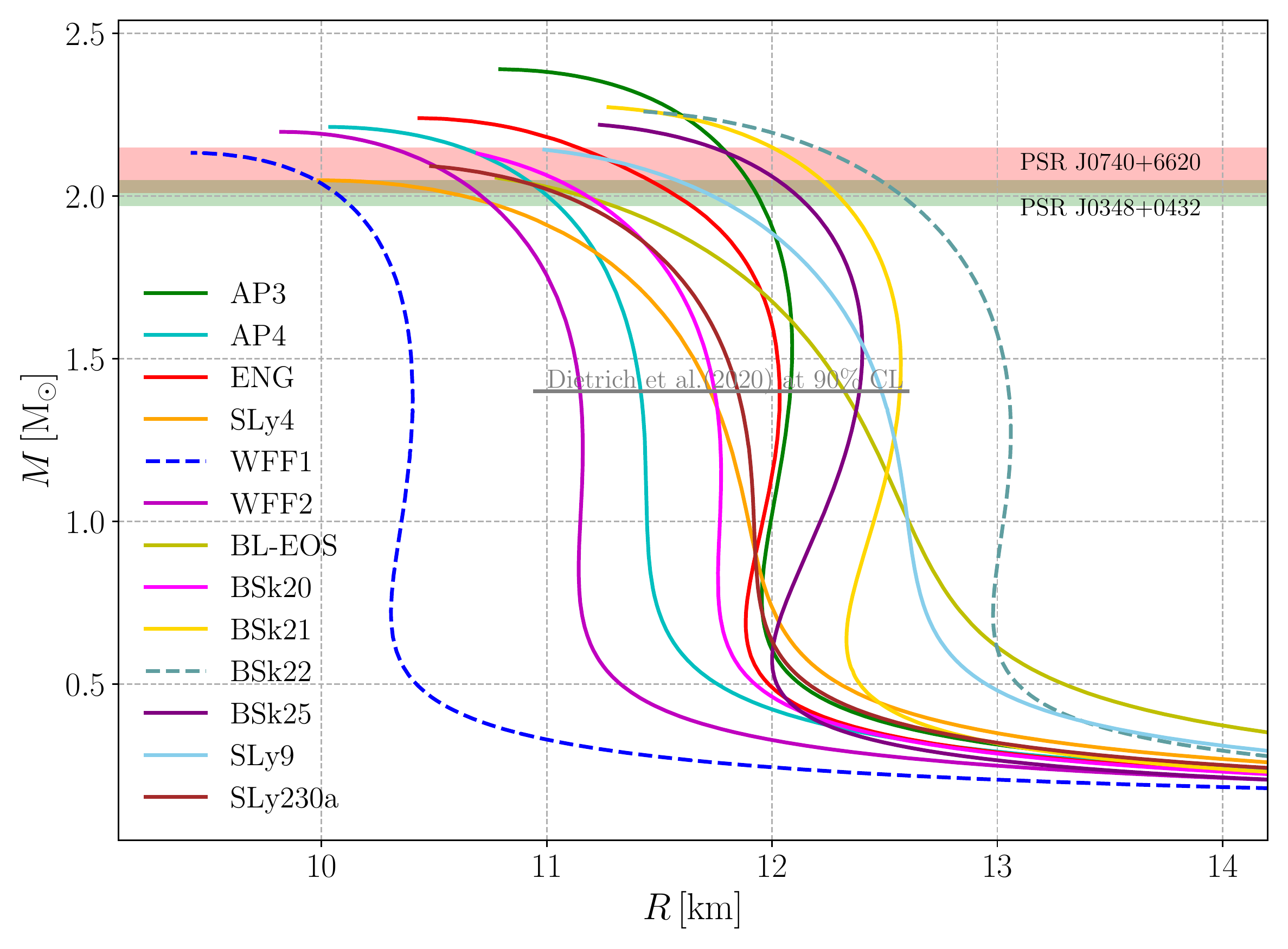}
\caption{The mass-radius relations of NSs in the
theory of GR for different EOSs we adopt. The solid lines represent EOSs that are consistent with the result
from Ref.~\cite{Dietrich:2020efo}, and the EOSs that deviate
are marked as dashed lines. The green and red regions depict the
 mass constraints with 1-$\sigma$ uncertainty from PSRs J0348$+$0432~\cite{Antoniadis:2013pzd} and
J0740$+$6620~\cite{Fonseca:2021wxt}, respectively.}
  \label{fig:MRRelation}
\end{figure*}

To derive the constraints on the DEF theory, we need to solve the structure of
NSs, which depend on the EOSs.  Refs.~\cite{Antoniadis:2013pzd,Fonseca:2021wxt}
indicate that the maximum mass of a NS should be somehow greater than
$2.0\,\Msun$. Therefore, we select 13 EOSs that are consistent with this
condition in our work. In addition, recently, Dietrich {\it et
al.}~\cite{Dietrich:2020efo} combined the multi-messenger observations of
GW170817, the X-ray and radio measurements of pulsars, and nuclear-theory
computations to constrain the NS's properties. As a result, they showed the
radius of a $1.4\,\Msun$ NS is $R = 11.75_{-0.81}^{+0.86} \, {\rm km}$ at the
90\% confidence level (CL); for other constraints, see
Ref.~\cite{LIGOScientific:2018cki} and references therein. The EOSs we use,
except for WFF1 and BSk22, are all supported by the results from
Ref.~\cite{Dietrich:2020efo}. The mass-radius relations are shown in
Fig.~\ref{fig:MRRelation} (see Ref.~\cite{Lattimer:2012nd} for a review).

Here, following the method
in Refs.~\cite{Shao:2017gwu,Zhao:2019suc,Guo:2021leu}, we apply MCMC techniques to obtain bounds on the spontaneous
scalarization. Given the priors, MCMC simulations evaluate the likelihood
functions $\sim 10^5$--$10^6$ times, and then derive the
posteriors that are consistent with observations. We use
the following logarithmic likelihood function,
\begin{equation}
  \hspace{-2.5cm} \ln \mathcal{L}_{\rm PSR} = 
    - \frac{1}{2} \sum_{i=1}^{N_{\rm PSR}} 
    \Bigg[ 
    \bigg( \frac{m_{\rm p} - m_{\rm p}^{\rm obs}}{\sigma_{m_{\rm p}^{\rm obs}}} \bigg)^2 + 
    \bigg( \frac{m_{\rm c} - m_{\rm c}^{\rm obs}}{\sigma_{m_{\rm c}^{\rm obs}}} \bigg)^2 +
    \bigg(\frac{ \dot P_{\rm b}^{\rm dipole} - 
    \dot P_{\rm b}^{\rm xs}}{\sigma_{\dot P_{\rm b}^{\rm xs}}} \bigg)^2 +
    \bigg( \frac{\gamma_{\rm E} - \gamma^{\rm obs}_{\rm E}}{\sigma_{\gamma^{\rm obs}_{\rm E}}} \bigg)^2 \Bigg]_i \,.
  \label{eq:lnL}
\end{equation}
Here, we use $N_{\rm PSR}$ binary pulsars, including their excess orbital decay
$\dot P_{\rm b}^{\rm xs}$, Einstein delay $\gamma^{\rm obs}_{\rm E}$, as well as
binary masses $m_{\rm p}^{\rm obs}$ and $m_{\rm c}^{\rm obs}$ ($\dot P_{\rm
b}^{\rm xs}$ is the observed orbital decay for each system $\dot P_{\rm b}^{\rm
obs}$ minus the sum of the kinematic effects caused by the Galactic acceleration
and the proper motion of the system and the prediction of GR for its orbital
decay~\cite{1970SvA....13..562S,Damour:1990wz}).  Note that the $\gamma_{\rm E}$
term is only included for PSR J0737$-$3039A~\cite{Kramer:2021jcw}. $\dot P_{\rm
b}^{\rm dipole}$ is the dipolar contribution to the orbital decay from the
scalar field. We use the DEF theory as an example. The prior
distribution for the theory parameter $\log_{10} \left| \alpha_0 \right|$ is
chosen to be flat in the range $[-5.3, -2.5]$. We choose the uniform prior on
the other theory parameter $-\beta_0$ in the range $[4.0, 4.8]$. The choice is
the same as Ref.~\cite{Zhao:2019suc}.

The usage of the ``observed'' masses from
Tables~\ref{tab:NSWD3s}--\ref{tab:DNS} in the combined test, i.e.\
Eq.~(\ref{eq:lnL}), requires some justification, since they were derived on the
basis of GR (or simply Newtonian gravity in case of the WD models).

WDs are weakly self-gravitating bodies ($\alpha_{\rm A} \simeq
\alpha_0$), and therefore masses obtained from spectroscopy or Shapiro delay are
practically identical to the GR masses. Differences are of the order
$\alpha_0^2$, which is globally constrained to $\lesssim 10^{-5}$ by the Cassini
experiment \cite{Damour:2007uf,Will:2018bme}, and to $\lesssim 10^{-7}$ by
pulsars in the $\beta_0$ range of interest here (see e.g.\
Ref.~\cite{Kramer:2021jcw}). Given that the mass ratio $q$ in
Table~\ref{tab:NSWD3s} is the same in GR and STG (see e.g.\
Ref.~\cite{Damour:2007uf}) and that the mass function (used to derive the pulsar
mass from the observed Shapiro shape parameter in Table~\ref{tab:extraNSWD}) is
very close to GR due to the small $\alpha_0$, we can safely use the GR pulsar
masses for all PSR-WD systems.

For the two PSR-NS systems in Table~\ref{tab:DNS}, the argument is less
obvious. In a single system test, we have explored the $\alpha_0$-$\beta_0$
space allowed by the observed post-Keplerian parameters published in
Refs.~\cite{Ferdman:2020huz,Kramer:2021jcw} (cf.\
Refs.~\cite{Damour:1996ke,Damour:1998jk} for the details of the method). For all
EOSs used here, we find that for $\beta_0 \le -4$ the masses determined by DEF
gravity deviate $\lesssim 2\times 10^{-3}\,\Msun$ for PSR~J1913$+$1102 and
$\lesssim 10^{-6}\,\Msun$ for PSR~J0737$-$3039 from those calculated with GR. In
both cases this is an order of magnitude smaller than the corresponding
(measurement) uncertainties.

To summarize, for the parameter space of interest in this paper, for all
seven systems the deviations from the GR masses are everywhere considerably
smaller than the corresponding mass uncertainties. Hence, the (GR) masses in
Tables~\ref{tab:NSWD3s}--\ref{tab:DNS} can be used directly in the combined
analysis.

%%%%%%%%%%%%%%%%%%%%%%%%%%%%%%%%%%%%%%%%%%%%%%%%%%%%%%%%%%%%%%%%%%%%%%%%%%%%%%%%

\section{Limits on scalar-tensor gravity}
\label{sec:limits}
%----------------------------------------------------------------------------------------%

In many cases, STG theories predict for the systems listed above a significant
scalar dipole and consequently a considerable loss of orbital energy via dipolar
GWs \cite{Will:2018bme}. For that reason, the close agreement of the intrinsic
$\dot{P}_{\rm b}$ with GR can be used to constrain these
theories~\cite{Wex:2014nva}. We focus on the DEF gravity as an example. In these
theories, the contribution of dipolar radiation to $\dot{P}_{\rm b}^{\rm xs}$
is, to leading order, given by
\begin{equation}\label{eq:PbdotD}
   \dot{P}_{\rm b}^{\rm dipole} = -\frac{4\pi^2G_\ast}{c^3P_{\rm b}}\,
                        \frac{m_{\rm p} m_{\rm c}}{m_{\rm p} + m_{\rm c}}\,
                        \frac{1 + e^2/2}{(1 - e^2)^{5/2}}\,
                        (\alpha_{\rm p} - \alpha_{\rm c})^2 \,,
\end{equation}
where $G_\ast=G_N/(1+\alpha_0^2)$ denotes the bare gravitational
constant \cite{Damour:1992we}, $\alpha_{\rm p}$ and $\alpha_{\rm c}$ are the
effective scalar couplings of the pulsar and the companion, and all other
relevant quantities can be found in Tables~\ref{tab:NSWD3s}--\ref{tab:DNS}.

\begin{figure*}
  \centering
  \includegraphics[width=\textwidth]{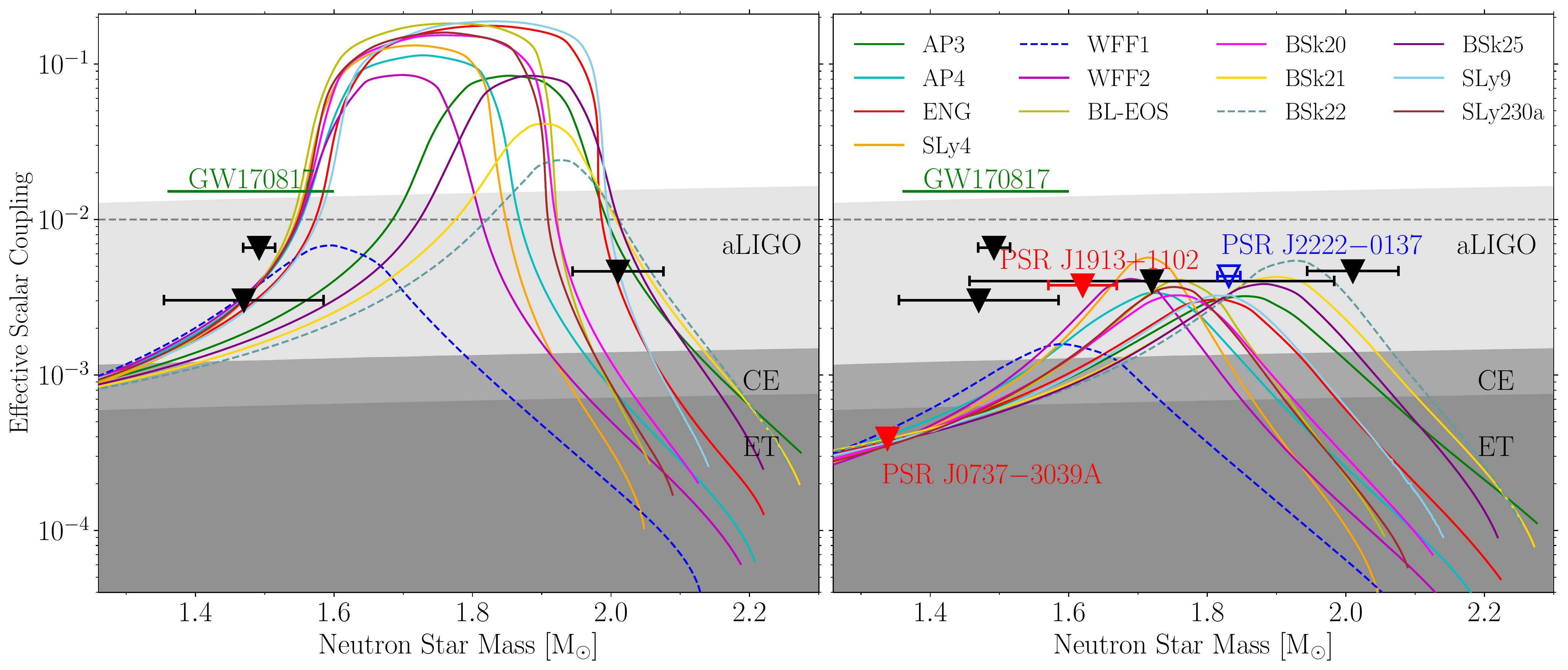}
\caption{\label{fig:before_after}The 90\% CL upper limits on the effective
scalar coupling $\alpha_{\rm A}$. On the left, we plot the limits obtained with the current experimental results for the pulsars considered by Ref.~\cite{Shao:2017gwu}. On the right, we see the limits obtained with all seven pulsars in our sample.}
\end{figure*}

The limits on $T_1(\alpha_0,\beta_0)$ determined using the above 7 binary
pulsars in a combined, multi-EOS, multi-pulsar analysis are shown in
Fig.~\ref{fig:before_after}. These cover the whole parameter space where
spontaneous scalarization can occur, and the full range of observed NS masses.
For a comparison, we repeat the same exercise without the recent results on the
intermediate-mass pulsars.

In this and the following figures, the black/red/blue triangles are the
upper limits at 90\% CL from individual pulsars. For parameters other
than $\dot P_{\rm b}^{\rm xs}$, we use their central values listed in
Tables~\ref{tab:NSWD3s}--\ref{tab:DNS}. For $\dot P_{\rm b}^{\rm xs}$, we
conservatively use a Gaussian function as its probability density function
(PDF),
\begin{equation}
  {\rm PDF \ of \ } {\dot P_{\rm b}^{\rm xs}} = \left\{ 
  \begin{array}{lr}
    {\cal N}\bigg(0, \big(\dot P_{\rm b}^{\rm xs}\big)^2 + \big( \sigma_{\dot P_{\rm b}^{\rm xs}} \big)^2 \bigg) ,                    & \dot P_{\rm b}^{\rm xs} > 0 \,, \\
    {\cal N}\bigg(\dot P_{\rm b}^{\rm xs},  \big( \sigma_{\dot P_{\rm b}^{\rm xs}} \big)^2 \bigg) ,                    & \dot P_{\rm b}^{\rm xs} \leq 0 \,.
  \end{array} \right.
\end{equation}
In addition, for five PSR-WD binaries, the triangles are the values
$\left(\alpha_{\rm p} - \alpha_{\rm c} \right)\approx \alpha_{\rm p}$. The
limits on $\alpha_{\rm A}$ from our combined analysis of binary pulsars
in Bayesian framework for different EOSs are shown as solid/dotted lines.

For comparison, we also show the result of the {\it difference} in effective
scalar coupling $\Delta \alpha$ from
GW170817~\cite{LIGOScientific:2018dkp}. In addition, for future GW
observatories, advanced LIGO (aLIGO) at its design sensitivity, Cosmic Explorer
(CE), and Einstein Telescope (ET), the expected limits on the $\Delta
\alpha$, i.e.\ the most optimistic limits on $\alpha_{\rm A}$, are given
via the method of Fisher information matrix. For them, a binary NS merger with a
1.25$\,\Msun$ companion at a luminosity distance of 200\,Mpc was assumed (see
e.g., Refs.~\cite{Shao:2017gwu, Zhao:2021bjw}).  Note that all upper limits on
the effective scalar couplings $\alpha_{\rm A}$, as well as uncertainties in the
NS masses are plotted at 90\% CL. The figure has two panels for easy comparison
of different pulsar sets.

Our main result comes out very clearly in Fig.~\ref{fig:before_after}:
the inclusion of the new systems with intermediate NS masses (PSRs~J1012+5307, 
J1913+1102, and J2222$-$0137), plus the precise results for PSR~J0737$-$3039A,
have greatly constrained spontaneous scalarization for the NS mass range where
it was still previously allowed, say in the range of $\left[ 1.50
\,\Msun,2.0\,\Msun \right]$.  Within this mass range, we find that $
\left|\alpha_{\rm A} \right| \gtrsim 0.006$ is excluded with significance larger than 90\%.
Furthermore, our analysis shows that although our limits on $\alpha_{\rm A}$
cannot be improved by the future advanced LIGO, even smaller deviations will be
detectable by the next generation of ground-based GW detectors (i.e., ET and
CE), which therefore have the potential to significantly improve on these new
limits from binary pulsars.

In order to understand the ability to constrain the parameter space from
different pulsars, we now investigate  limits on $\alpha_{\rm A}$  using
different sets, listed in Table~\ref{tab:pulsar_sets}.

\begin{table}[!htbp]
  \caption{Different scenarios for investigating the upper limits on $\alpha_{\rm A}$.\label{tab:pulsar_sets}}
\begin{indented}
    \lineup
    \item[]\begin{tabular}{l|ccccccc}  \br
      Scenarios & J0348 & J1738 & J1909 & J1012 & J2222 & J1913 &
J0737 \\ \mr
\textsf{NSWD3s} & \checkmark & \checkmark & \checkmark & ... & ... & ... & ... \\ 
\textsf{NSWD4s} & \checkmark & \checkmark & \checkmark & \checkmark & ... & ... & ... \\ 
\textsf{NSWD5s} & \checkmark & \checkmark & \checkmark & \checkmark & \checkmark & ... & ... \\ 
\textsf{NSWD5s+J1913} & \checkmark & \checkmark & \checkmark & \checkmark &
\checkmark & \checkmark & ... \\ 
\textsf{J0737} & ... & ... & ... & ... & ... & ... & \checkmark \\ 
\textsf{NSWD5s+J1913+J0737} & \checkmark & \checkmark & \checkmark & \checkmark
& \checkmark &  \checkmark & \checkmark \\ \br
  \end{tabular}
\end{indented}
\end{table}

\begin{figure*}
  \centering
  \includegraphics[width=\textwidth]{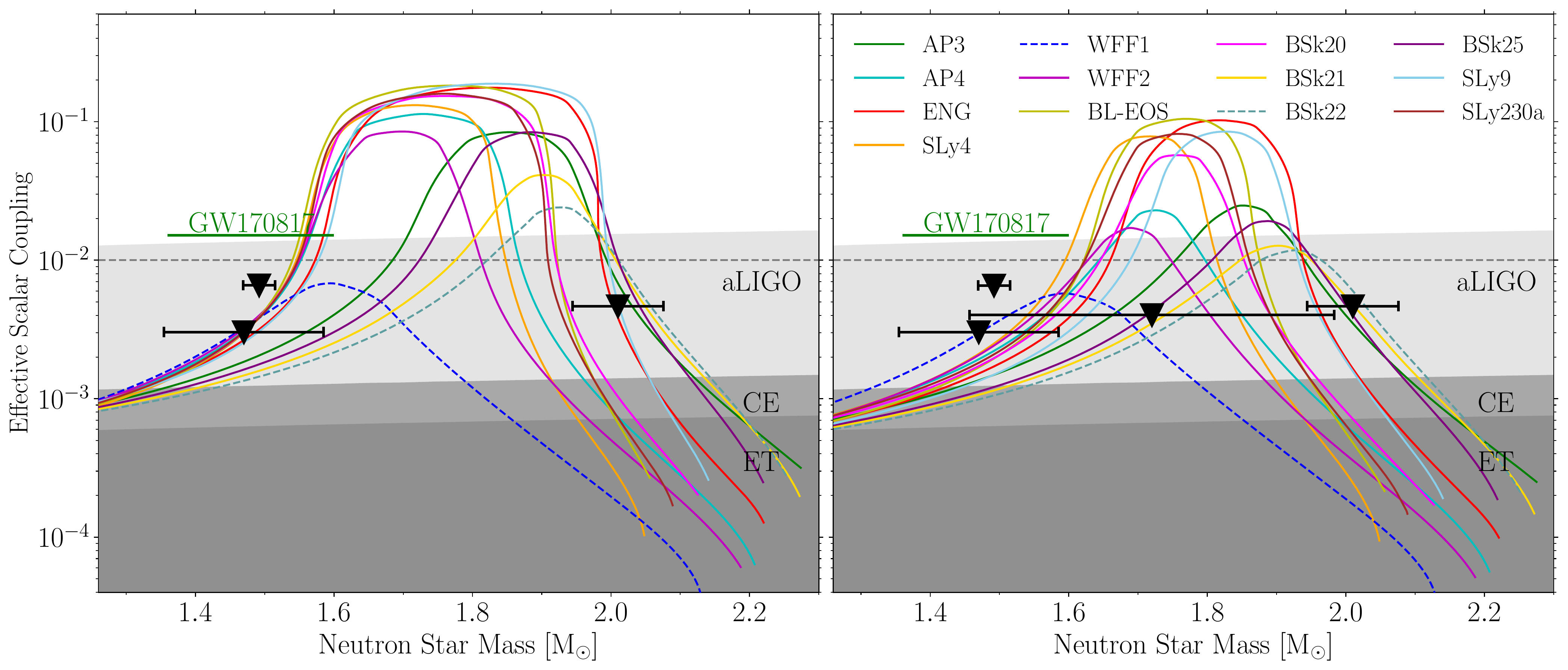}
\caption{\label{fig:NSWD3s_vs_NSWD4s} Same as Fig.~\ref{fig:before_after} but for the \textsf{NSWD3s} (left) and
\textsf{NSWD4s} (right) binary sets.}
\end{figure*}

\begin{figure*}
  \centering
  \includegraphics[width=\textwidth]{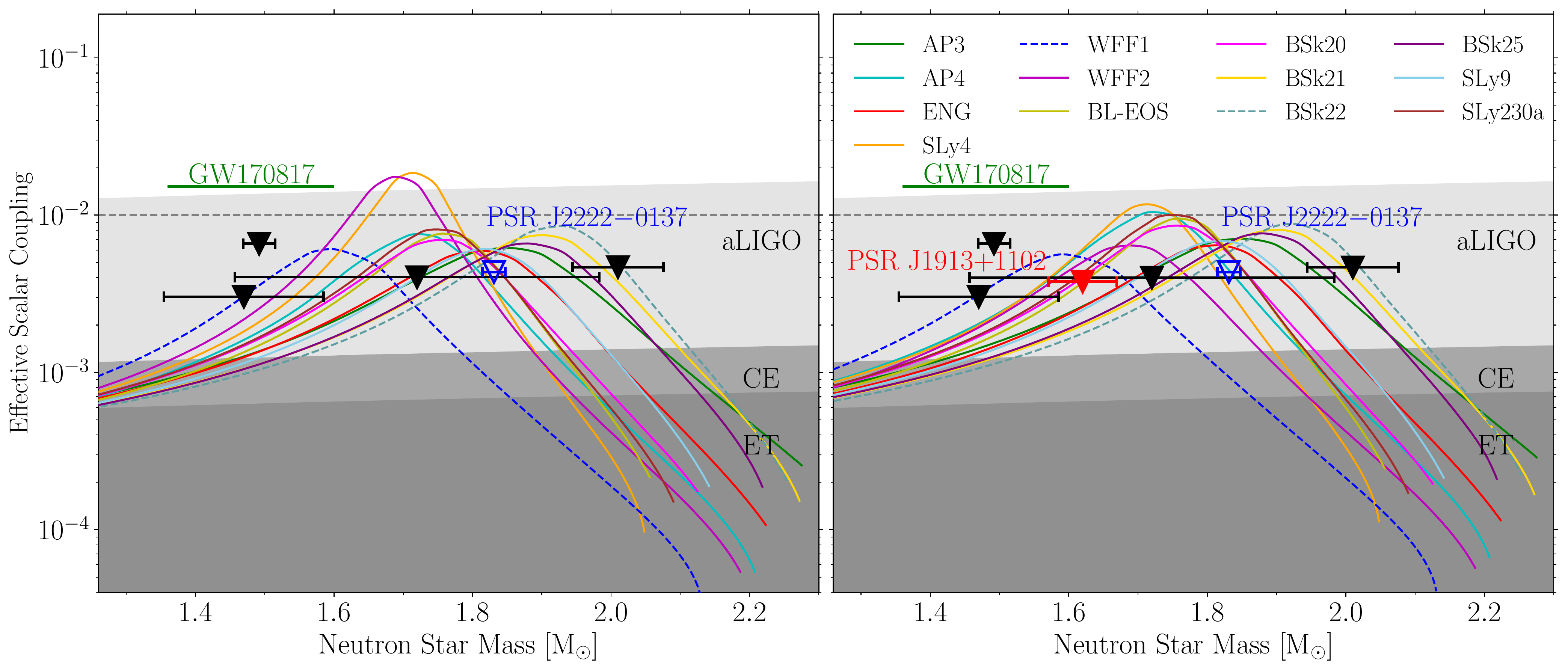}
\caption{\label{fig:NSWD5s_vs_NSWD6s} Same as Fig.~\ref{fig:before_after},
but for the \textsf{NSWD5s} (left) and \textsf{NSWD5s+J1913} (right) binary sets.}
\end{figure*}

\begin{figure*}
  \centering
  \includegraphics[width=\textwidth]{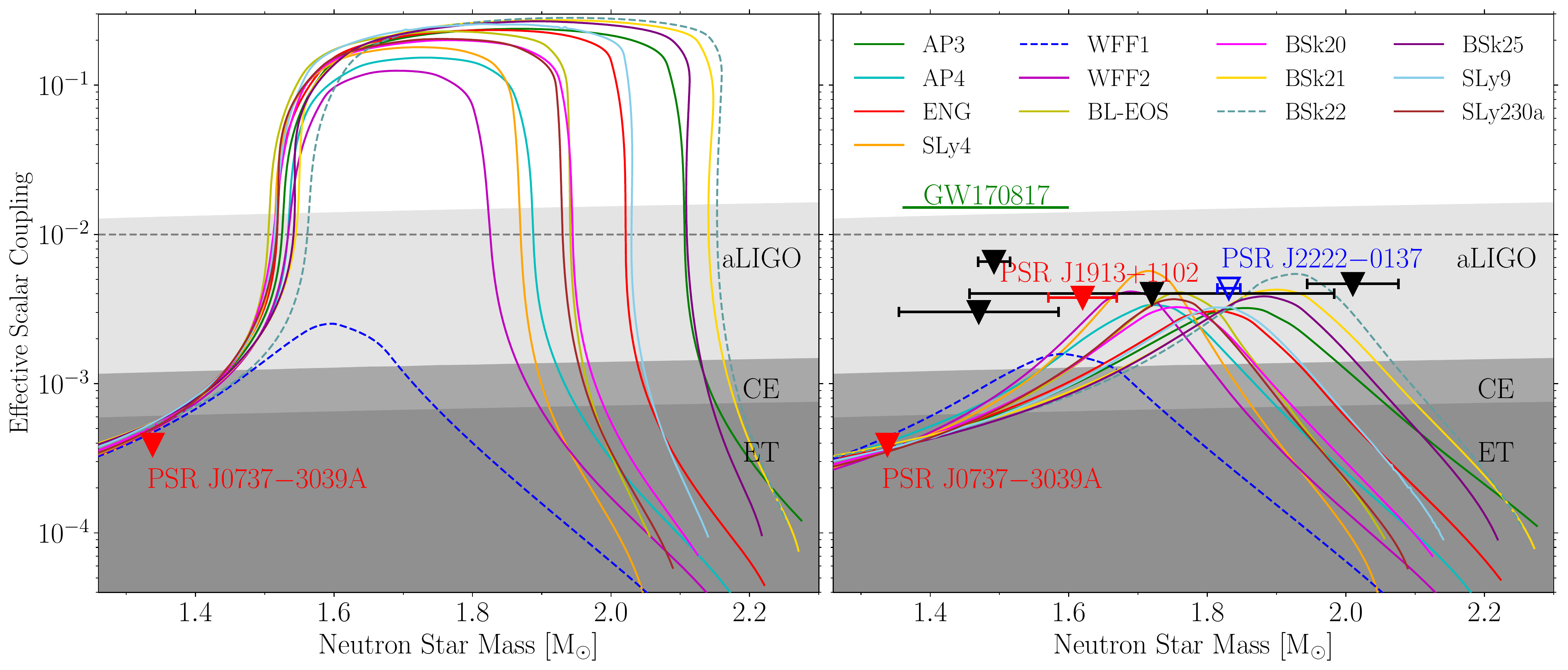}
\caption{\label{fig:J0737_vs_NSWD7s} Same as Fig.~\ref{fig:before_after},
but for the \textsf{J0737} (left) and
\textsf{NSWD5s+J1913+J0737} (right) binary sets.}
\end{figure*}

For each of those sets, we performed the MCMC technique for all the EOSs we
adopted, as described above. The results are shown in
Figs.~\ref{fig:NSWD3s_vs_NSWD4s}, \ref{fig:NSWD5s_vs_NSWD6s}
and~\ref{fig:J0737_vs_NSWD7s},

In Fig.~\ref{fig:NSWD3s_vs_NSWD4s}, the left panel shows the constraints at 90\%
CL with only PSRs~J0348+0432, J1738+0333, and J1909$-$3744 (same as the
left panel in Fig.~\ref{fig:before_after}), the right panel shows the effect of
introducing PSR~J1012+5307. As we can see, for all the 13 EOSs we
consider~\cite{Lattimer:2000nx}, the inclusion of PSR~J1012+5307 narrows down
the mass range of spontaneous scalarization and improves the upper limits on
$\alpha_{\rm A}$ to $ \left|\alpha_{\rm A} \right| \lesssim 0.1$. The individual bound from
PSR~J1012+5307 is competitive, but due to the large uncertainty of its
mass, it introduces only weak constraints to the global upper limits on
$\alpha_{\rm A}$.  {Therefore, a wide $m_A \in \left[ 1.60
\,\Msun,1.95\,\Msun \right]$ scalarization window remains open.}

Figure~\ref{fig:NSWD5s_vs_NSWD6s} further introduces two binaries, PSRs
J2222$-$0137 and J1913$+$1102, which are shown in the left and right panels,
respectively. With the help of these binaries, on the whole mass range,
$\left|\alpha_{\rm A}\right|$ has been constrained to be $\lesssim 10^{-2}$ with
90\% confidence, which roughly equals to the (optimistic) limit expected from
future aLIGO for a binary NS merger at a luminosity distance of $200 \, {\rm
Mpc}$ \cite{KAGRA:2013pob}. The significantly improved constraints result
directly from these two pulsars having well determined masses that are well
within the mass gap identified in Fig.~\ref{fig:NSWD3s_vs_NSWD4s}. These
two pulsars contribute fundamentally to the closure of the gap of mass
scalarization.

After performing Bayesian analysis for \textsf{J0737} (see the left panel in
Fig.~\ref{fig:J0737_vs_NSWD7s}), we find that, for the soft EOS WFF1,
PSR~J0737$-$3039A puts the tightest bound on $\alpha_{\rm A}$. The reason is
that  NS masses of  PSR J0737$-$3039 are close to the scalarization window of
WFF1. However, for EOSs with higher scalarization masses,
PSR~J0737$-$3039A provides by itself no significant constraints, owing to the
low masses of the two NSs in that system.  Despite that, even for those EOSs,
the inclusion of PSR~J0737$-$3039A in the joint analysis with all other systems
is still useful, with lower limits on $\alpha_{\rm A}$ for the whole NS mass
interval, as we can see from a comparison of the right panel (the same
as in Fig.~\ref{fig:before_after}) with the right panel of
Fig.~\ref{fig:NSWD5s_vs_NSWD6s}. This improvement results from a
tighter limit on $\alpha_0$ from the inclusion of the extremely precise $\dot
P_{\rm b}$ of PSR~J0737$-$3039A.

%%%%%%%%%%%%%%%%%%%%%%%%%%%%%%%%%%%%%%%%%%%%%%%%%%%%%%%%%%%%%%%%%%%%%%%%%%%%%%%%

\section{Conclusions}
\label{sec:conclusion}

In this paper we have used 7 binary pulsars in total to further constrain the
NSs' effective scalar coupling ($\alpha_{\rm A}$) possible within the DEF
gravity, in particular for values of $\beta_0$ between $-4.8$ and $-4.0$ where
such theories predict strong-field spontaneous scalarization,
leading to large scalar charges. In this region, and by extension
in the whole parameter space of DEF gravity, we find $\left|\alpha_{\rm
A}\right| \lesssim 0.006$. This excludes spontaneous scalarization for the
observed range of NS masses, and applies to the full range (from
soft to stiff) of allowed EOSs.\footnote{We have ignored more exotic
possibilities, like phase transitions (see e.g.\ Ref.~\cite{Blaschke:2019tbh}).
Anyway, excluding certain fine-tuned scenarios, we do not expect a significant
change to the constraints of DEF gravity presented here, given the dense
coverage on NS masses used in our combined test.}

These conclusions confirm, in their broad lines, a recent
analysis~\cite{Chiba:2021rqa}, but do so much more thoroughly. That work reaches
the conclusion that the scalarization gap is closed based solely on the new
experimental results of PSR~J1012+5307 and especially PSR~J2222$-$0137, and by
analyzing a limited set of EOSs. Furthermore, in that study the effective scalar
couplings are obtained without considering the uncertainties on the masses of
the components of these binaries. If we take these uncertainties into account,
within a wider set of EOSs, we can see (in the left panel of
Fig.~\ref{fig:NSWD5s_vs_NSWD6s}) that, for some EOSs, a limited amount of
spontaneous scalarization is still possible for NS masses around $1.7 \, \Msun$.
It is only by including the recent results on two PSR-NS systems
(PSRs~J0737$-$3039A and J1913+1102) that we can close this gap for all NS masses
and EOSs considered in this work. Nevertheless, the analysis of
Ref.~\cite{Chiba:2021rqa} is in its general aspects significant, particularly in
the clear identification of the importance of the new experimental results of
PSR~J2222$-$0137~\cite{Guo:2021bqa}.

More generally, the analysis presented here demonstrates the absence
of (significant) dipolar radiation in binary pulsars for NS masses up to $\sim
2\,\Msun$. This should be of relevance for other gravity theories as well, in
particular for theories that predict a similar mass-dependent dynamics for binary
pulsars. To give an example, for sufficiently negative $\beta_0$, the
Mendes-Ortiz (MO) gravity \cite{Mendes:2016fby} also shows a strongly non-linear
behavior (accompanied with very large scalar charges) in the mass range
investigated here \cite{Anderson:2019hio}.

We do not want to leave it unmentioned that there are various ways to modify
GR such that one can have strongly scalarized NSs ($\alpha_{\rm A} \sim 1$) and,
at the same time, avoid the limits obtained here. For instance, as already
emphasized by Antoniadis {\it et al.}~\cite{Antoniadis:2013pzd}, binary pulsar
experiments are naturally insensitive to sufficiently short-range/massive
fields. They can therefore not exclude spontaneous-scalarization scenarios like
in massive DEF gravity \cite{Ramazanoglu:2016kul, Yazadjiev:2016pcb,
Xu:2020vbs}, just to give an example.\footnote{See, however, the discussion in
Refs.~\cite{Hu:2020ubl, Hu:2021tyw} on potential future pulsar tests to
constrain certain short-range modifications of GR that evade current pulsar
experiments.} While the limits presented in Fig.~\ref{fig:before_after} are
already below what is expected for aLIGO, for short-range gravity phenomena
related to NSs there is certainly a valuable complementarity between pulsar
observations and GW astronomy (see also the discussions on dynamical
scalarization in Ref.~\cite{Shao:2017gwu} and on finite-range scalar forces in
Ref.~\cite{Sagunski:2017nzb}).  Seen in a broader context, our best knowledge of
gravity requires the combination of all possible experiments, and pulsar tests
that cover such a wide range of NS masses certainly play an important role in
this.

Finally, looking into the future, there are good prospects for a
relatively fast improvement in the precision of the radiative tests with these
binary pulsars, not only from the increase in their timing baselines, but also
from the inclusion of new telescopes, like the FAST
telescope~\cite{Nan:2011um,2016RaSc...51.1060L,Li:2019ads} and MeerKAT
array~\cite{Bailes:2020qai,Kramer:2021xvm}, which are already taking data of
much higher quality than what was possible until now. These will further improve
with the Square Kilometre Array~\cite{Shao:2014wja, Weltman:2018zrl}. These
improved limits will either tighten further the limits on dipolar GW emission,
or lead to their detection and the falsification of GR.

%%%%%%%%%%%%%%%%%%%%%%%%%%%%%%%%%%%%%%%%%%%%%%%%%%%%%%%%%%%%%%%%%%%%%%%%%%%%%%%%

\section*{Acknowledgements}

We thank Robert D.\ Ferdman for discussions, and Alexander Batrakov,
Daniela D.\ Doneva, Fethi M.\ Ramazano\u{g}lu, Luciano Rezzolla, Thomas Tauris,
Stoytcho S.\ Yazadjiev, and anonymous referees for helpful comments.  This work
was supported by the National SKA Program of China (2020SKA0120300), the
National Natural Science Foundation of China (12147177, 11975027, 11991053,
11721303), the Young Elite Scientists Sponsorship Program by the China
Association for Science and Technology (2018QNRC001), the Max Planck Partner
Group Program funded by the Max Planck Society, and the High-performance
Computing Platform of Peking University. J. Zhao was supported by the ``LiYun''
postdoctoral fellowship of Beijing Normal University.


\section*{References}

\bibliographystyle{iopart-num}
\bibliography{refs} % if your bibtex file is called example.bib

\providecommand{\newblock}{}
\begin{thebibliography}{10}
\expandafter\ifx\csname url\endcsname\relax
  \def\url#1{{\tt #1}}\fi
\expandafter\ifx\csname urlprefix\endcsname\relax\def\urlprefix{URL }\fi
\providecommand{\eprint}[2][]{\url{#2}}

\bibitem{Einstein:1915ca}
Einstein A 1915 {\em Sitzungsber. Preuss. Akad. Wiss. Berlin (Math. Phys.)\/}
  {\bf 1915} 844--847

\bibitem{Einstein:1918btx}
Einstein A 1918 {\em Sitzungsber. Preuss. Akad. Wiss. Berlin (Math. Phys. )\/}
  {\bf 1918} 154--167

\bibitem{Peters:1964zz}
Peters P~C 1964 {\em Phys. Rev.\/} {\bf 136} B1224--B1232

\bibitem{Hulse:1974eb}
Hulse R~A and Taylor J~H 1975 {\em Astrophys. J. Lett.\/} {\bf 195} L51--L53

\bibitem{Taylor:1979zz}
Taylor J~H, Fowler L~A and McCulloch P~M 1979 {\em Nature\/} {\bf 277} 437--440

\bibitem{Weisberg:2016jye}
Weisberg J~M and Huang Y 2016 {\em Astrophys. J.\/} {\bf 829} 55
  (\textit{Preprint} \eprint{1606.02744})

\bibitem{Damour:2014tpa}
Damour T 2015 {\em Class. Quant. Grav.\/} {\bf 32} 124009 (\textit{Preprint}
  \eprint{1411.3930})

\bibitem{Fujii:2003pa}
Fujii Y and Maeda K 2007 {\em {The Scalar-tensor Theory of Gravitation}\/}
  Cambridge Monographs on Mathematical Physics (Cambridge, England: Cambridge
  University Press)

\bibitem{Damour:1993hw}
Damour T and Esposito-Far\`ese G 1993 {\em Phys. Rev. Lett.\/} {\bf 70}
  2220--2223

\bibitem{Jordan:1949zz}
Jordan P 1949 {\em Nature\/} {\bf 164} 637--640

\bibitem{Jordan:1959eg}
Jordan P 1959 {\em Z. Phys.\/} {\bf 157} 112--121

\bibitem{Fierz:1956zz}
Fierz M 1956 {\em Helv. Phys. Acta\/} {\bf 29} 128--134

\bibitem{Brans:1961sx}
Brans C and Dicke R~H 1961 {\em Phys. Rev.\/} {\bf 124} 925--935

\bibitem{Mirshekari:2013vb}
Mirshekari S and Will C~M 2013 {\em Phys. Rev. D\/} {\bf 87} 084070
  (\textit{Preprint} \eprint{1301.4680})

\bibitem{Yagi:2021loe}
Yagi K and Stepniczka M 2021 {\em Phys. Rev. D\/} {\bf 104} 044017
  (\textit{Preprint} \eprint{2105.01614})

\bibitem{Abbott:2018lct}
Abbott B {\em et~al.\/} (LIGO Scientific and Virgo Collaborations) 2019 {\em
  Phys. Rev. Lett.\/} {\bf 123} 011102 (\textit{Preprint} \eprint{1811.00364})

\bibitem{Zhao:2019suc}
Zhao J, Shao L, Cao Z and Ma B~Q 2019 {\em Phys. Rev. D\/} {\bf 100} 064034
  (\textit{Preprint} \eprint{1907.00780})

\bibitem{Niu:2021nic}
Niu R, Zhang X, Wang B and Zhao W 2021 {\em Astrophys. J.\/} {\bf 921} 149
  (\textit{Preprint} \eprint{2105.13644})

\bibitem{TheLIGOScientific:2017qsa}
Abbott B {\em et~al.\/} (LIGO Scientific and Virgo Collaborations) 2017 {\em
  Phys. Rev. Lett.\/} {\bf 119} 161101 (\textit{Preprint} \eprint{1710.05832})

\bibitem{LIGOScientific:2020ibl}
Abbott R {\em et~al.\/} (LIGO Scientific, Virgo) 2021 {\em Phys. Rev. X\/} {\bf
  11} 021053 (\textit{Preprint} \eprint{2010.14527})

\bibitem{LIGOScientific:2021qlt}
Abbott R {\em et~al.\/} (LIGO Scientific, KAGRA, VIRGO) 2021 {\em Astrophys. J.
  Lett.\/} {\bf 915} L5 (\textit{Preprint} \eprint{2106.15163})

\bibitem{Damour:1996ke}
Damour T and Esposito-Far\`ese G 1996 {\em Phys. Rev. D\/} {\bf 54} 1474--1491
  (\textit{Preprint} \eprint{gr-qc/9602056})

\bibitem{Shao:2017gwu}
Shao L, Sennett N, Buonanno A, Kramer M and Wex N 2017 {\em Phys. Rev. X\/}
  {\bf 7} 041025 (\textit{Preprint} \eprint{1704.07561})

\bibitem{Anderson:2019eay}
Anderson D, Freire P and Yunes N 2019 {\em Class. Quant. Grav.\/} {\bf 36}
  225009 (\textit{Preprint} \eprint{1901.00938})

\bibitem{Antoniadis:2012vy}
Antoniadis J, van Kerkwijk M~H, Koester D, Freire P~C~C, Wex N, Tauris T~M,
  Kramer M and Bassa C~G 2012 {\em Mon. Not. Roy. Astron. Soc.\/} {\bf 423}
  3316 (\textit{Preprint} \eprint{1204.3948})

\bibitem{Freire:2012mg}
Freire P~C~C, Wex N, Esposito-Farese G, Verbiest J~P~W, Bailes M, Jacoby B~A,
  Kramer M, Stairs I~H, Antoniadis J and Janssen G~H 2012 {\em Mon. Not. Roy.
  Astron. Soc.\/} {\bf 423} 3328 (\textit{Preprint} \eprint{1205.1450})

\bibitem{Antoniadis:2013pzd}
Antoniadis J {\em et~al.\/} 2013 {\em Science\/} {\bf 340} 6131
  (\textit{Preprint} \eprint{1304.6875})

\bibitem{Wex:2014nva}
Wex N 2014 {Testing Relativistic Gravity with Radio Pulsars} {\em {Frontiers in
  Relativistic Celestial Mechanics: Applications and Experiments}\/} vol~2 ed
  Kopeikin S~M (Walter de Gruyter GmbH, Berlin/Boston) p~39 (\textit{Preprint}
  \eprint{1402.5594})

\bibitem{Shao:2016ezh}
Shao L and Wex N 2016 {\em Sci. China Phys. Mech. Astron.\/} {\bf 59} 699501
  (\textit{Preprint} \eprint{1604.03662})

\bibitem{Shibata:2013pra}
Shibata M, Taniguchi K, Okawa H and Buonanno A 2014 {\em Phys. Rev. D\/} {\bf
  89} 084005 (\textit{Preprint} \eprint{1310.0627})

\bibitem{LIGOScientific:2017vwq}
Abbott B~P {\em et~al.\/} (LIGO Scientific, Virgo) 2017 {\em Phys. Rev.
  Lett.\/} {\bf 119} 161101 (\textit{Preprint} \eprint{1710.05832})

\bibitem{Lazaridis:2009kq}
Lazaridis K {\em et~al.\/} 2009 {\em Mon. Not. R. Astron. Soc.\/} {\bf 400}
  805--814 (\textit{Preprint} \eprint{0908.0285})

\bibitem{Antoniadis:2016hxz}
Antoniadis J, Tauris T~M, Ozel F, Barr E, Champion D~J and Freire P~C~C 2016
  (\textit{Preprint} \eprint{1605.01665})

\bibitem{Desvignes:2016yex}
Desvignes G {\em et~al.\/} 2016 {\em Mon. Not. Roy. Astron. Soc.\/} {\bf 458}
  3341--3380 (\textit{Preprint} \eprint{1602.08511})

\bibitem{Reardon:2015kba}
Reardon D~J {\em et~al.\/} 2016 {\em Mon. Not. Roy. Astron. Soc.\/} {\bf 455}
  1751--1769 (\textit{Preprint} \eprint{1510.04434})

\bibitem{NANOGrav:2017wvv}
Arzoumanian Z {\em et~al.\/} (NANOGrav) 2018 {\em Astrophys. J. Suppl.\/} {\bf
  235} 37 (\textit{Preprint} \eprint{1801.01837})

\bibitem{Cognard:2017xyr}
Cognard I {\em et~al.\/} 2017 {\em Astrophys. J.\/} {\bf 844} 128
  (\textit{Preprint} \eprint{1706.08060})

\bibitem{MataSanchez:2020pys}
Mata~S\'anchez D, Istrate A~G, van Kerkwijk M~H, Breton R~P and Kaplan D~L 2020
  {\em Mon. Not. Roy. Astron. Soc.\/} {\bf 494} 4031--4042 (\textit{Preprint}
  \eprint{2004.02901})

\bibitem{Ding:2020sig}
Ding H, Deller A~T, Freire P, Kaplan D~L, Lazio T~J~W, Shannon R and Stappers B
  2020 {\em Astrophys. J.\/} {\bf 896} 85 [Erratum: Astrophys.J. 900, 89
  (2020)] (\textit{Preprint} \eprint{2004.14668})

\bibitem{Liu:2020hkx}
Liu K {\em et~al.\/} 2020 {\em Mon. Not. Roy. Astron. Soc.\/} {\bf 499}
  2276--2291 (\textit{Preprint} \eprint{2009.12544})

\bibitem{Guo:2021bqa}
Guo Y~J {\em et~al.\/} 2021 {\em Astron. Astrophys.\/} {\bf 654} A16
  (\textit{Preprint} \eprint{2107.09474})

\bibitem{Lazarus:2016hfu}
Lazarus P {\em et~al.\/} 2016 {\em Astrophys. J.\/} {\bf 831} 150
  (\textit{Preprint} \eprint{1608.08211})

\bibitem{Ferdman:2020huz}
Ferdman R~D {\em et~al.\/} 2020 {\em Nature\/} {\bf 583} 211--214
  (\textit{Preprint} \eprint{2007.04175})

\bibitem{Kramer:2021jcw}
Kramer M {\em et~al.\/} 2021 {\em Phys. Rev. X\/} {\bf 11} 041050
  (\textit{Preprint} \eprint{2112.06795})

\bibitem{Kramer:2006nb}
Kramer M {\em et~al.\/} 2006 {\em Science\/} {\bf 314} 97--102
  (\textit{Preprint} \eprint{astro-ph/0609417})

\bibitem{Fonseca:2021wxt}
Fonseca E {\em et~al.\/} 2021 {\em Astrophys. J. Lett.\/} {\bf 915} L12
  (\textit{Preprint} \eprint{2104.00880})

\bibitem{Miller:2021qha}
Miller M~C {\em et~al.\/} 2021 {\em Astrophys. J. Lett.\/} {\bf 918} L28
  (\textit{Preprint} \eprint{2105.06979})

\bibitem{Dietrich:2020efo}
Dietrich T, Coughlin M~W, Pang P~T~H, Bulla M, Heinzel J, Issa L, Tews I and
  Antier S 2020 {\em Science\/} {\bf 370} 1450--1453 (\textit{Preprint}
  \eprint{2002.11355})

\bibitem{LIGOScientific:2018cki}
Abbott B~P {\em et~al.\/} (LIGO Scientific, Virgo) 2018 {\em Phys. Rev.
  Lett.\/} {\bf 121} 161101 (\textit{Preprint} \eprint{1805.11581})

\bibitem{Lattimer:2012nd}
Lattimer J~M 2012 {\em Ann. Rev. Nucl. Part. Sci.\/} {\bf 62} 485--515
  (\textit{Preprint} \eprint{1305.3510})

\bibitem{Guo:2021leu}
Guo M, Zhao J and Shao L 2021 {\em Phys. Rev. D\/} {\bf 104} 104065
  (\textit{Preprint} \eprint{2106.01622})

\bibitem{1970SvA....13..562S}
{Shklovskii} I~S 1970 {\em Soviet Astronomy\/} {\bf 13} 562

\bibitem{Damour:1990wz}
Damour T and Taylor J~H 1991 {\em Astrophys. J.\/} {\bf 366} 501--511

\bibitem{Damour:2007uf}
Damour T 2007 {Binary Systems as Test-beds of Gravity Theories} {\em {6th
  SIGRAV Graduate School in Contemporary Relativity and Gravitational Physics:
  A Century from Einstein Relativity: Probing Gravity Theories in Binary
  Systems}\/} (\textit{Preprint} \eprint{0704.0749})

\bibitem{Will:2018bme}
Will C~M 2018 {\em {Theory and Experiment in Gravitational Physics}\/}
  (Cambridge University Press)

\bibitem{Damour:1998jk}
Damour T and Esposito-Farese G 1998 {\em Phys. Rev. D\/} {\bf 58} 042001
  (\textit{Preprint} \eprint{gr-qc/9803031})

\bibitem{Damour:1992we}
Damour T and Esposito-Far\`ese G 1992 {\em Class. Quant. Grav.\/} {\bf 9}
  2093--2176

\bibitem{LIGOScientific:2018dkp}
Abbott B~P {\em et~al.\/} (LIGO Scientific, Virgo) 2019 {\em Phys. Rev.
  Lett.\/} {\bf 123} 011102 (\textit{Preprint} \eprint{1811.00364})

\bibitem{Zhao:2021bjw}
Zhao J, Shao L, Gao Y, Liu C, Cao Z and Ma B~Q 2021 {\em Phys. Rev. D\/} {\bf
  104} 084008 (\textit{Preprint} \eprint{2106.04883})

\bibitem{Lattimer:2000nx}
Lattimer J~M and Prakash M 2001 {\em Astrophys. J.\/} {\bf 550} 426
  (\textit{Preprint} \eprint{astro-ph/0002232})

\bibitem{KAGRA:2013pob}
Abbott B~P {\em et~al.\/} (KAGRA, LIGO Scientific, VIRGO) 2018 {\em Living Rev.
  Rel.\/} {\bf 21} 3 (\textit{Preprint} \eprint{1304.0670})

\bibitem{Blaschke:2019tbh}
Blaschke D, Alvarez-Castillo D~E, Ayriyan A, Grigorian H, Largani N~K and Weber
  F 2020 {\em {Astrophysical aspects of general relativistic mass twin
  stars}\/} (WORLD SCIENTIFIC) chap Chapter 7, pp 207--256 (\textit{Preprint}
  \eprint{1906.02522})

\bibitem{Chiba:2021rqa}
Chiba T 2021 {\em Prog. Theor. Exp. Phys.\/} (\textit{Preprint}
  \eprint{2104.11362}) \urlprefix\url{https://doi.org/10.1093/ptep/ptab138}

\bibitem{Mendes:2016fby}
Mendes R~F~P and Ortiz N 2016 {\em Phys. Rev. D\/} {\bf 93} 124035
  (\textit{Preprint} \eprint{1604.04175})

\bibitem{Anderson:2019hio}
Anderson D and Yunes N 2019 {\em Class. Quant. Grav.\/} {\bf 36} 165003
  (\textit{Preprint} \eprint{1901.00937})

\bibitem{Ramazanoglu:2016kul}
Ramazano\u{g}lu F~M and Pretorius F 2016 {\em Phys. Rev. D\/} {\bf 93} 064005
  (\textit{Preprint} \eprint{1601.07475})

\bibitem{Yazadjiev:2016pcb}
Yazadjiev S~S, Doneva D~D and Popchev D 2016 {\em Phys. Rev. D\/} {\bf 93}
  084038 (\textit{Preprint} \eprint{1602.04766})

\bibitem{Xu:2020vbs}
Xu R, Gao Y and Shao L 2020 {\em Phys. Rev. D\/} {\bf 102} 064057
  (\textit{Preprint} \eprint{2007.10080})

\bibitem{Hu:2020ubl}
Hu H, Kramer M, Wex N, Champion D~J and Kehl M~S 2020 {\em Mon. Not. Roy.
  Astron. Soc.\/} {\bf 497} 3118--3130 (\textit{Preprint} \eprint{2007.07725})

\bibitem{Hu:2021tyw}
Hu Z, Gao Y, Xu R and Shao L 2021 {\em Phys. Rev. D\/} {\bf 104} 104014
  (\textit{Preprint} \eprint{2109.13453})

\bibitem{Sagunski:2017nzb}
Sagunski L, Zhang J, Johnson M~C, Lehner L, Sakellariadou M, Liebling S~L,
  Palenzuela C and Neilsen D 2018 {\em Phys. Rev. D\/} {\bf 97} 064016
  (\textit{Preprint} \eprint{1709.06634})

\bibitem{Nan:2011um}
Nan R, Li D, Jin C, Wang Q, Zhu L, Zhu W, Zhang H, Yue Y and Qian L 2011 {\em
  Int. J. Mod. Phys. D\/} {\bf 20} 989--1024 (\textit{Preprint}
  \eprint{1105.3794})

\bibitem{2016RaSc...51.1060L}
{Li} D and {Pan} Z 2016 {\em Radio Science\/} {\bf 51} 1060--1064
  (\textit{Preprint} \eprint{1612.09372})

\bibitem{Li:2019ads}
{Li} D, {Dickey} J~M and {Liu} S 2019 {\em Research in Astronomy and
  Astrophysics\/} {\bf 19} 016 (\textit{Preprint} \eprint{1904.05882})

\bibitem{Bailes:2020qai}
Bailes M {\em et~al.\/} 2020 {\em Publ. Astron. Soc. Austral.\/} {\bf 37} e028
  (\textit{Preprint} \eprint{2005.14366})

\bibitem{Kramer:2021xvm}
Kramer M {\em et~al.\/} 2021 {\em Mon. Not. Roy. Astron. Soc.\/} {\bf 504}
  2094--2114 (\textit{Preprint} \eprint{2102.05160})

\bibitem{Shao:2014wja}
Shao L {\em et~al.\/} 2015 {\em PoS\/} {\bf AASKA14} 042 (\textit{Preprint}
  \eprint{1501.00058})

\bibitem{Weltman:2018zrl}
Weltman A {\em et~al.\/} 2020 {\em Publ. Astron. Soc. Austral.\/} {\bf 37} e002
  (\textit{Preprint} \eprint{1810.02680})

\end{thebibliography}

\end{document}